\documentclass[%
 aip,jcp,
 amsmath,amssymb,
 reprint,
]{revtex4-2}

\usepackage[a-1b]{pdfx} 

\usepackage{graphicx}
\usepackage{dcolumn}
\usepackage{bm}

\usepackage[utf8]{inputenc}
\usepackage[T1]{fontenc}
\usepackage{mathptmx} 
\usepackage{listings}
\usepackage{rotating} 

\usepackage{color}
\usepackage{dcolumn} 
\usepackage{bm} 
\usepackage{graphicx}
\usepackage{multirow} 
\usepackage{pifont} 
\usepackage{epsfig}
\usepackage{amsmath} 
\usepackage{subfigure}
\usepackage{float}
\usepackage{booktabs}
\usepackage{tabularx}
\usepackage{natbib}
\usepackage{gensymb}
\usepackage{rotating}
\usepackage{threeparttable}
\usepackage{comment}
\usepackage[normalem]{ulem}

\begin{document}
  
\author{A. Eugene DePrince III}
\email{adeprince@fsu.edu}
\affiliation{
             Department of Chemistry and Biochemistry,
             Florida State University,
             Tallahassee, FL 32306-4390, USA}

\author{Stephen H. Yuwono}
\affiliation{
             Department of Chemistry and Biochemistry,
             Florida State University,
             Tallahassee, FL 32306-4390, USA}
             
\title{Static Electric Dipole Polarizability and Hyperpolarizability Tensors from Mean-Field Cavity Quantum Electrodynamics Approaches}


\begin{abstract}
First-order electric dipole response functions are implemented for cavity quantum electrodynamics (QED) generalizations of Hartree-Fock (HF) and Kohn-Sham density functional theory (DFT) in order to assess the degree to which static molecular response properties are impacted by interactions between electronic degrees of freedom and an optical cavity mode.  Isotropically averaged static electric dipole polarizability tensors from QED-HF and QED-DFT are found to be somewhat insensitive to the presence of the cavity under realistic single-molecule coupling strengths. In contrast, the first hyperpolarizability tensor computed using QED-HF can be significantly modified by the cavity, depending on the relative orientation of the molecule and cavity mode polarization axis. For example, compared to the isolated molecule case, the isotropically averaged hyperpolarizability for $p$-nitroaniline decreases by more than 20\% when the molecule is coupled to a single-mode optical cavity with a coupling strength of $|{\bm \lambda}| = 0.05$ a.u. and when the cavity mode is polarized along the principal molecular axis. On the other hand, with this coupling strength and polarization, the isotropically averaged static dipole polarizability from QED-HF or QED-DFT decreases by only $\approx$ 2--5 \%, depending on the choice of DFT functional.  
\end{abstract}

\maketitle

\section{Introduction}

The last decade has seen an explosion in interest in multicomponent quantum chemical theories that treat photon and electron degrees of freedom on equal footing.\cite{DePrince23_041301} These theoretical developments have been driven by numerous experiments demonstrating that strong and ultra-strong coupling between light and matter degrees of freedom can be leveraged for energy\cite{Lidzey14_712,Sanvitto17_e16212,Schwartz18_105,Forrest20_2002127,Lidzey21_16661,GomezRivas22_123,Musser22_2105569,Schwartz23_338} or charge\cite{Giebink20_177401,Ebbesen20_10219,George21_13616} transport applications, to augment materials properties,\cite{Whittaker98_6697,Mugnier04_036404,Bramati07_106401,Ebbesen15_1123,Ebbesen16_2403,Bellessa19_173902,Ebbesen16_7352,KenaCohen18_119,Singer19_1801682,Ebbesen21_1486,Singer21_085307,Giebink22_7937,Borjesson18_2273,Borjesson21_3255} or even as a means to control chemical transformations.\cite{Ebbesen12_1592, Ebbesen16_11462, Shegai18_eaas9552, Ebbesen19_615, Ebbesen19_15324, George19_10635, Uji-i20_5332, Ebbesen20_10436, Borjesson21_2010737, Ebbesen21_16877, Shalabney21_chemrxiv.7234721.v5, George21_379, Ebbesen21_5712, George22_195, Jeffrey22_429} A large swath of the quantum chemical toolkit has been extended to model such applications within the framework of cavity quantum electrodynamics (QED), including mean-field approximations  \cite{Bauer11_042107,Rubio14_012508,Rubio15_093001,Rubio18_992,Appel19_225,Rubio19_2757,Narang20_094116,Rubio22_7817, Rubio23_11191,Narang23_383,DePrince22_9303,Rubio22_094101,DePrince23_5264, DePrince24_064109,Bauer11_042107,Rubio14_012508,Tokatly13_233001,Rubio17_3026,Tokatly18_235123,Varga22_194106,Shao21_064107, Shao22_124104, DePrince22_9303, DePrince23_5264,Foley22_154103,Wilson24_094111}  single-reference many-body wave function approaches,\cite{Narang23_arXiv:2307.14822, Reichman24_1143,Koch20_041043, Manby20_023262, Corni21_6664, DePrince21_094112, Flick21_9100, Koch21_094113, Koch22_234103, Flick22_4995, Rubio22_094101, Knowles22_204119, DePrince22_054105, Rubio23_2766, Rubio23_10184, Koch23_4938, Koch23_8988, Koch23_031002, DePrince23_5264,DePrince24_064109,Koch24_8876,Stopkowicz24_9572,Koch24_8838,Koch24_e1684,Koch25_021040,Dreuw23_124128} and multireference theories.\cite{DePrince22_053710,Yu24_032804,Foley24_1214,Foley25_chemrxiv-2025-q6rfm,Ronca25_6862,Huo23_2353,Huo24_16184,Foley24_174105}


Many studies applying {\em ab initio} QED methods have focused on the ground-state problem, with a common question coming to the forefront: can electronic strong coupling lead to non-trivial changes in ground-state electronic structure that can be leveraged for useful purposes in chemistry applications? Along these lines, many exciting predictions have been made, including enhanced diastereocontrol,\cite{DePrince22_9303} pronounced changes to reaction barrier heights,\cite{Flick22_4995,Rubio23_2766,Rubio23_10184} precise control over reaction products,\cite{Rubio23_2766,Rubio23_10184} modified ionization potentials and electron affinities,\cite{DePrince21_094112,Koch22_234103} and selective (de-)stabilization of non-covalently bound complexes.\cite{Koch21_094113} 
So, within the limits of the established models, the answer to this question appears to be yes, with the caveat that it will be difficult to verify many of these predictions experimentally, as simulations often involve coupling strengths that are larger than what has been experimentally achieved.\cite{Flick22_4995,Rubio23_2766,Rubio23_10184,DePrince24_064109} 
 

With the exception of linear absorption spectra, somewhat less attention has been paid to the problem of {\em ab initio} response property prediction, particularly for nonlinear response properties, in cavity-embedded systems. This is somewhat surprising, given that numerous experiments have indicated that strong coupling leads to a variety of interesting nonlinear phenomena, including enhanced second-\cite{Ebbesen16_7352} or third-harmonic\cite{KenaCohen18_119,Singer19_1801682,Ebbesen21_1486} generation, novel four-wave mixing processes\cite{Bramati07_106401} and modulated electroabsorption responses.\cite{Giebink22_7937} Nonetheless, only a handful of studies have explored {\em ab initio} molecular response property prediction in cavity QED environments. In Ref.~\citenum{Koch24_e1684}, Castagnola, Riso, Ronca, Barlini, and Koch present the framework of exact and approximate polaritonic response theories, as well as some numerical data for linear response properties (QED Hartree-Fock [QED-HF] oscillator strengths plus transition properties related to the photon degrees of freedom). In Ref.~\citenum{Koch24_7841}, Barlini, Bianchi, Ronca, and Koch develop linear response theory involving static magnetic fields and nuclear spin degrees of freedom and present data corresponding to QED-HF derived  magnetizabilities and nucleus independent chemical shifts. Schnappinger and Kowalewski\cite{Kowalewski25_5171} recently explored cavity-induced changes to the static electric dipole polarizability under vibrational strong coupling at the QED-HF level (within the cavity Born-Oppenheimer approximation). In Ref.~\citenum{Koch25_4447}, Castagnola, Riso, El Moutaoukal, Ronka, and Koch extended QED-HF response theory to the strong-coupling QED-HF approach\cite{Koch22_1368,Koch24_8911} and evaluated oscillator strengths and photon transition moments in the length and velocity gauges. 
As far as we can tell, there has yet to be a numerical study exploring cavity effects on higher-than-linear molecular response properties within an {\em ab initio} response theory framework. That said, Narang and coworkers have applied real-time (RT) quantum electrodynamical density functional theory (DFT) to the description of second-,\cite{Narang23_383,Narang24_369} third-,\cite{Narang23_383} and fourth-harmonic generation.\cite{Narang24_369} 
RT electronic structure methods\cite{Lopata20_9951} have long been applied to the simulation of non-linear optical responses,\cite{Saalfrank07_034107,Rehr07_154114,HeadGordon12_909,Madsen13_062511,Li13_064104,Madsen14_063416,Repisky16_5823,Luppi17_947} and the RT formulation of the problem is particularly desirable in the context of DFT-based methods because it avoids the need to evaluate high-order derivatives of the exchange--correlation functional.

In this work, we develop and apply a response theory formalism for mean-field {\em ab initio} QED approaches. In Sec.~\ref{SEC:THEORY}, we derive static first-order response functions for QED-HF and QED-DFT, which allow us to compute the static electric dipole polarizability tensor ($\alpha$) for cavity-embedded systems. For QED-HF, we also use the first-order response function to evaluate the the first hyperpolarizability tensor ($\beta$). Section \ref{SEC:THEORY} provides expressions for these quantities, as well as the relevant details of the underlying QED-HF and QED-DFT approaches. We then study the effects that strong coupling to an optical cavity mode have on these first- and second-order properties. The details of the computational aspects of this study are presented in Sec.~\ref{SEC:COMPUTATIONAL_DETAILS}, and the numerical results of our simulations are presented in Sec.~\ref{SEC:RESULTS}. Some concluding remarks can be found in Sec.~\ref{SEC:CONCLUSIONS}.



\section{Theory}

\label{SEC:THEORY}

Interactions between electronic degrees of freedom and a single optical cavity mode can be modeled using the Pauli-Fierz Hamiltonian,\cite{Spohn04_book,Rubio18_0118} which we represent within the dipole approximation and in the length gauge as
\vspace{-0.1cm}
\begin{align}
        \label{EQN:PFH}
\hat{H}_\text{PF}
    &= \hat{H}_\text{e} + \omega_\text{cav} \hat{b}^\dagger\hat{b} - \sqrt{\frac{\omega_\text{cav}}{2}} {\bm \lambda}\cdot \hat{\bm{\mu}} \left(\hat{b}^\dagger + \hat{b}\right) \nonumber \\
    &+ \frac{1}{2} \left({\bm \lambda}\cdot \hat{\bm{\mu}}\right)^2
\end{align}
Here, $\hat{H}_{\rm e}$ is the usual electronic Hamiltonian, the second term represents the Hamiltonian for the cavity mode, which is characterized by a frequency $\omega_\text{cav}$ and coupling vector ${\bm \lambda}$. The operators $\hat{b}^\dagger$ and $\hat{b}$ are bosonic creation and annihilation operators that create or destroy one photon in the photon mode, respectively. The third term represents dipolar coupling between the electron and photon degrees of freedom, and the last term is the molecular dipole self-energy (DSE). The symbol $\hat{{\bm \mu}}=\hat{{\bm \mu}}_{\rm e} + \hat{{\bm \mu}}_{\rm nu}$ represents the total dipole operator for the molecular component of the system. It will also be convenient to introduce the coherent-state (CS) representation of the Hamiltonian,
\begin{align}
\hat{H}_\text{CS} &= \hat{U}_\text{CS}^\dagger \hat{H}_\text{PF} \hat{U}_\text{CS} \nonumber \\
        \label{EQN:PFH_COHERENT}
     &= \hat{H}_\text{e} + \omega_\text{cav} \hat{b}^\dagger\hat{b} - \sqrt{\frac{\omega_\text{cav}}{2}} {\bm \lambda}\cdot [\hat{\bm{\mu}}_\text{e} - \langle \hat{\bm{\mu}}_\text{e}\rangle ] \left(\hat{b}^\dagger + \hat{b}\right) \nonumber \\
    &+ \frac{1}{2} \left({\bm \lambda}\cdot [\hat{\bm{\mu}}_\text{e} - \langle \hat{\bm{\mu}}_\text{e}\rangle ]\right)^2
\end{align}
where the operator 
\begin{equation}
\label{EQN:U_CS}
    \hat{U}_{\rm CS} = {\rm exp}\left( \frac{-{\bm \lambda} \cdot \langle {\hat{\bm{\mu}}} \rangle }{\sqrt{2 \omega_{\rm cav}}} \left (\hat{b}^{\dagger} - \hat{b} \right ) \right) 
\end{equation} 
is the CS transformation operator that diagonalizes the photon part of $\hat{H}_\text{PF}$.\cite{Koch20_041043}

\subsection{QED-HF and QED-DFT}

The simplest approximate eigenfunction of $\hat{H}_\text{PF}$ is the QED-HF state
\begin{equation}
\label{EQN:QED_HF}
    |\Phi_{0}\rangle = |0_\text{e}\rangle \hat{U}_\text{CS}|0_{\rm p}\rangle
\end{equation}
which is a direct product of a Slater determinant of electronic orbitals ($|0_\text{e}\rangle$) and a zero-photon state ($\hat{U}_\text{CS}|0_{\rm p}\rangle$), where $|0_{\rm p}\rangle$ represents the photon vacuum. Taking the expectation value of $\hat{H}_\text{PF}$ with respect to this function and integrating out the photon degrees of freedom gives the QED-HF energy 
\begin{align}
    E_\text{0} &= \langle 0_\text{e}| \langle 0_\text{p}| \hat{U}^\dagger_\text{CS} \hat{H}_\text{PF} \hat{U}_\text{CS} | 0_\text{p} \rangle |0_\text{e}\rangle \nonumber \\
    &= \langle 0_\text{e}| \hat{H}_\text{CS} |0_\text{e}\rangle \nonumber \\
    \label{EQN:QED_HF_ENERGY}
    &= \langle 0_\text{e}| \hat{H}_\text{e} |0_\text{e}\rangle + \frac{1}{2} \langle 0_\text{e} | \left({\bm \lambda}\cdot[\hat{\bm{\mu}}_\text{e} - \langle \hat{\bm{\mu}}_\text{e}\rangle ]\right)^2 |0_\text{e}\rangle
\end{align}
where the expectation value of the dipole operator is taken with respect to $|0_\text{e}\rangle$. Note that only the electronic part of the dipole operator appears in this expression.
Because there are no photon degrees of freedom in Eq.~\ref{EQN:QED_HF_ENERGY}, $|0_\text{e}\rangle$ can be determined from a standard self-consistent field (SCF) procedure, with modified one- and two-electron integrals that account for the presence of the DSE term. 

From here, one can adapt QED-HF to obtain a cavity QED version of Kohn-Sham DFT, or QED-DFT.\cite{DePrince22_9303,Rubio22_094101,DePrince23_5264,DePrince24_064109} The QED-DFT ground-state is taken to be the non-interacting state in Eq.~\ref{EQN:QED_HF}, but with $|0_\text{e}\rangle$ now representing a determinant of Kohn-Sham orbitals. The electronic part of the energy, $\langle 0_\text{e}| \hat{H}_\text{e} |0_\text{e}\rangle$, is handled as in standard Kohn-Sham DFT, {\em i.e.}, it is given by the sum of contributions from the core Hamiltonian and classical Coulomb contributions, plus an exchange-correlation functional that depends only on the electronic degrees of freedom. The DSE term is treated in the same way as in QED-HF theory. In this work, no explicit electron--photon exchange--correlation (XC) functional is employed, but several such functionals have been developed and applied elsewhere.\cite{Rubio15_093001,Rubio17_113036,Rubio18_992,Rubio21_e2110464118,Tokatly23_235424,Rubio24_052823,Flick25_073002}

\subsection{Mean-Field Cavity QED Response Theory}

We define the Hamiltonian for a molecule embedded within a single-mode optical cavity and perturbed by a static electric field (${\bm \epsilon}$) as
\begin{align}
\label{EQN:PERTURBED_H}
    \hat{H} = \hat{H}_\text{PF} - \sum_\alpha \hat{\mu}_{\text{e},\alpha} \epsilon_\alpha
\end{align}
where $\hat{\mu}_{\text{e},\alpha}$ refers to the $\alpha$-component of the electric dipole operator. The QED-HF wave function for the perturbed system is 
\begin{align}
    |\Phi({\bm \kappa})\rangle = \text{exp}(\hat{\kappa})|\Phi_0\rangle
\end{align}
where $\hat{\kappa}$, which encodes the electronic and photonic responses of the system to the applied field, is defined by $\hat{\kappa} = \hat{\kappa}_\text{e} + \hat{\kappa}_\text{p}$ with
\begin{align}
\label{EQN:KAPPA_E}
    \hat{\kappa}_\text{e} &= \sum_{ia} \left ( \kappa_{\text{e},ia} \hat{a}^\dagger_a \hat{a}_i - \kappa_{\text{e},ai}^* \hat{a}^\dagger_i \hat{a}_a \right )\\
\label{EQN:KAPPA_P}
    \hat{\kappa}_\text{p} &= \kappa_\text{p} \hat{b}^\dagger - \kappa_\text{p}^* \hat{b}
\end{align}
Here, $\hat{a}^\dagger_i$ ($\hat{a}_i$) refers to a fermionic creation (annihilation) operator for orbital $i$. Throughout this work, the labels $i$, $j$, and $k$ and $a$, $b$, and $c$ refer to molecular spin orbitals that are occupied or unoccupied in $|0_\text{e}\rangle$, respectively. In Eq.~\ref{EQN:PERTURBED_WFN}, ${\bm \kappa}$ (without the hat) represents a vector containing the electronic and photonic amplitudes appearing in Eqs.~\ref{EQN:KAPPA_E} and \ref{EQN:KAPPA_P}. For real-valued orbitals and static applied electric fields, $\kappa_{\text{e},ia} = \kappa_{\text{e},ai}^*$ and $\kappa_{\text{p}} = \kappa_{\text{p}}^*$, so $\hat{U}_\text{CS}$ and $\text{exp}(\hat{\kappa})$ commute, and we can write
\begin{align}
\label{EQN:PERTURBED_WFN}
    |\Phi({\bm \kappa})\rangle = \hat{U}_\text{CS}~ \text{exp} (\hat{\kappa}) |0_\text{e} \rangle|0_\text{p}\rangle
\end{align}

The energy of the perturbed system is given by the expectation value
\begin{align}
\label{EQN:PERTURBED_ENERGY}
    E &= \langle 0_\text{p} |  \langle 0_\text{e}|\text{exp}(-\hat{\kappa}) \hat{H}_\text{CS} \text{exp} (\hat{\kappa}) |0_\text{e} \rangle|0_\text{p}\rangle \nonumber \\
    &- \sum_\alpha \epsilon_\alpha \langle 0_\text{e} | \text{exp}(-\hat{\kappa}_\text{e}) \hat{\mu}_{\text{e},\alpha} \text{exp} (\hat{\kappa}_\text{e}) |0_\text{e} \rangle  
\end{align}
The energy can also be expanded as a Taylor series in the applied field, as
\begin{align}
\label{EQN:ENERGY_RESPONSE}
E &= E_0 - \sum_\alpha  \mu_{\text{e},\alpha} \epsilon_\alpha - \frac{1}{2} \sum_{\alpha\beta}  \alpha_{\text{e},\alpha\beta} \epsilon_\alpha \epsilon_\beta \nonumber \\ 
  &- \frac{1}{6} \sum_{\alpha\beta\gamma}  \beta_{\text{e},\alpha\beta\gamma} \epsilon_\alpha \epsilon_\beta\epsilon_\gamma  + \ldots
\end{align}
where $E_0$ represents the energy of the unperturbed system, and the elements of the electric dipole ($\mu_{\text{e},\alpha}$), dipole polarizability ($\alpha_{\text{e},\alpha\beta}$), and first hyperpolarizability ($\beta_{\text{e},\alpha\beta\gamma}$) tensors are defined by various derivatives with respect to ${\bm \epsilon}$, {\em i.e.}
\begin{align}
\mu_{\text{e},\alpha} &= -\left . \frac{d E} {d \epsilon_\alpha} \right |_{{\bm \epsilon} = 0}   \\
\alpha_{\text{e},\alpha\beta} &= -\left . \frac{d^2 E} {d \epsilon_\alpha d \epsilon_\beta} \right |_{{\bm \epsilon} = 0} \\
\beta_{\text{e},\alpha\beta\gamma} &= -\left . \frac{d^3 E} {d \epsilon_\alpha d \epsilon_\beta d \epsilon_\gamma} \right |_{{\bm \epsilon} = 0}
\end{align}
By applying differential operators
\begin{align}
    \frac{d}{d\epsilon_\alpha} = \frac{\partial}{\partial \epsilon_\alpha} + \sum_{ia} \frac{\partial \kappa_{\text{e}, ia}}{\partial \epsilon_\alpha} \frac{\partial}{\partial \kappa_{\text{e}, ia}} + \frac{\partial \kappa_{\text{p}}}{\partial \epsilon_\alpha} \frac{\partial}{\partial \kappa_\text{p}}
\end{align}
to Eq.~\ref{EQN:PERTURBED_ENERGY}, we can obtain expressions for these response properties. Doing so, the QED-HF electric dipole
is
\begin{align}
\label{EQN:QED_HF_DIPOLE}
    \mu_{\text{e},\alpha} &= \left . \langle \Phi({\bm \kappa}) | \hat{\mu}_{\text{e},\alpha}|\Phi({\bm \kappa})\rangle\right |_{\bm \epsilon = 0} \nonumber \\
    &- \sum_{ia} \left . \langle \Phi({\bm \kappa}) | [\hat{H}, \hat{\tau}_{ai}]|\Phi({\bm \kappa})\rangle\right |_{\bm \epsilon = 0} \nonumber \\
    &- \left . \langle \Phi({\bm \kappa}) | [\hat{H}, \hat{b}^\dagger - \hat{b}]|\Phi({\bm \kappa})\rangle\right |_{\bm \epsilon = 0} 
\end{align}
with $\hat{\tau}_{ai} = \hat{a}^\dagger_a \hat{a}_i - \hat{a}^\dagger_i \hat{a}_a$. The second term is the orbital gradient, which is zero when evaluated at zero field, and the third term also evaluates to zero, so the electric dipole moment is simply
\begin{align}
    \mu_{\text{e},\alpha} = \langle 0_\text{e} | \hat{\mu}_{\text{e},\alpha}|0_\text{e}\rangle
\end{align}

Repeating this exercise for the electric dipole polarizability leads to the same form as in standard HF theory, {\em i.e.},
\begin{align}
    \alpha_{\text{e},\alpha\beta} =  2\sum_{ia} \mu_{\alpha,ia} \kappa_{\text{e},ia}^\beta
\end{align}
where
\begin{align}
    \mu_{\alpha, ia} = \langle 0_\text{e} | \hat{\mu}_{\text{e},\alpha}\hat{a}^\dagger_a \hat{a}_i | 0_\text{e}\rangle 
\end{align}
and 
\begin{align}
    \kappa_{\text{e},ia}^\beta = \left . \frac{\partial \kappa_{\text{e}, ia}}{\partial \epsilon_\beta} \right |_{\bm \epsilon = 0}
\end{align}
A key difference from standard HF theory, though, is that this first-order electronic response vector will also depend on the first-order response of the photonic part of the system. Equations for these first-order responses can be obtained by differentiating the stationary conditions (the commutators that evaluate to zero in Eq.~\ref{EQN:QED_HF_DIPOLE}) with respect to the field. Doing so leads to 
\begin{align}
\sum_{jb} (A_{ia, jb} + B_{ia, jb}) \kappa^\alpha_{\text{e}, ia} - \sqrt{2 \omega_\text{cav}}\mu^\alpha_{ia}  \kappa_\text{p}^\alpha = \mu_{ia}^\alpha  \\ 
-\sqrt{2\omega_\text{cav}} \sum_{ia}  \mu_{ia}^\alpha \kappa_{\text{e}, ia} ^\alpha + \kappa_\text{p}^\alpha \omega_\text{cav} = 0
\end{align}
where ${\bf A}$ and ${\bf B}$ are the usual matrices that arise in standard time-dependent HF theory / the random-phase approximation (RPA), augmented by DSE contributions (see Ref.~\citenum{DePrince22_9303} for explicit expressions). Here, we have also introduced the shorthand notation
\begin{align}
    \kappa_\text{p}^\alpha = \left . \frac{\partial \kappa_\text{p}}{\partial \epsilon_\alpha} \right |_{{\bm \epsilon} = 0}
\end{align}
Note that first-order response vectors for QED-DFT can be obtained from the same equations by modifying the RPA matrices to include the appropriate derivatives with respect to the XC functional. 

Knowledge of the first-order response of the system is sufficient to determine second-order properties. For QED-HF, the expression for the first hyperpolarizability has a similar form as the expression for standard HF theory\cite{Ruud06_1}
\begin{align}
    \beta_{\text{e},\alpha\beta\gamma} &= -\sum_{ia,jb,kc} \kappa^\alpha_{\text {e}, jb} \kappa^\beta_{\text {e}, ia} \kappa^{\gamma}_{\text{e},kc} E^{[3]}_{ia,jb,kc} \nonumber \\
    & - \sum_{ia,jb} \left [\kappa^\alpha_{\text {e}, jb} \kappa^\beta_{\text {e}, ia} h^{\gamma}_{ia,jb} + \kappa^\alpha_{\text {e}, jb} \kappa^\gamma_{\text {e}, ia} h^{\beta}_{ia,jb} + \kappa^\beta_{\text {e}, jb} \kappa^\gamma_{\text {e}, ia} h^{\alpha}_{ia,jb}\right ]
\end{align}
Here, $E^{[3]}$ represents the third derivative of the energy with respect to the electronic part of ${\bm \kappa}$
\begin{align}
    E^{[3]}_{ia,jb,kc} =\left . \langle \Phi({\bm \kappa})| [[[\hat{H}, \hat{\tau}_{ai}],\hat{\tau}_{bj}],\hat{\tau}_{ck}]|\Phi({\bm \kappa})\rangle \right |_{{\bm \epsilon} = 0}
\end{align}
The symbol 
$h^\alpha_{ia,jb}$ represents a combination of two terms: (i) the first derivative of the energy with respect to the electric field in the $\alpha$ direction and the second derivative with respect to the electronic part of ${\bm \kappa}$ and (ii) a mixed third derivative of the energy with respect to the electronic and photonic parts of ${\bm \kappa}$,
\begin{align}
    &h^{\alpha}_{ia,jb} = -\left .\langle \Phi({\bm \kappa})| [[\hat{\mu}_{\text{e},\alpha}, \hat{\tau}_{ai}],\hat{\tau}_{bj}]|\Phi({\bm \kappa})\rangle \right |_{{\bm \epsilon} = 0} \nonumber \\
    &+ \kappa^\alpha_{\text{p}} \left .\langle \Phi({\bm \kappa})| [[[\hat{H}, \hat{\tau}_{ai}],\hat{\tau}_{bj}],\hat{b}^\dagger - \hat{b}]|\Phi({\bm \kappa})\rangle \right |_{{\bm \epsilon} = 0}
\end{align}

\section{Computational Details}

\label{SEC:COMPUTATIONAL_DETAILS}

The first-order QED-HF and QED-DFT response equations were implemented in \texttt{hilbert},\cite{hilbert} which is a plugin to the \textsc{Psi4} electronic structure package.\cite{Sherrill20_184108} 
Expressions for the QED-HF and QED-DFT static electric dipole polarizabilities and QED-HF static first hyperpolarizabilities were implemented in the same plugin. The correctness of the non-QED HF and DFT static polarizability tensors and the non-QED HF hyperpolarizability tensor were verified numerically against the implementation in Gaussian.\cite{g16} For the QED versions of these methods, the correctness of the implementation was verified numerically against finite-field calculations. 

In Sec.~\ref{SEC:RESULTS}, we explore the effect of an optical cavity on static response properties for $p$-nitroaniline. Molecular geometries for this molecule were obtained from optimizations carried out at the standard (non-QED) HF and DFT levels, using the d-aug-cc-pVTZ basis set.\cite{Dunning89_1007,Harrison92_6796,Dunning94_2975} Separate DFT optimizations were carried out using the SVWN,\cite{Slater74_book,Nusair80_1200} PBE,\cite{Ernzerhof96_3865,Ernzerhof97_1396} B3LYP,\cite{Parr88_785,Becke93_5648,Frisch94_11623} and $\omega$B97X\cite{Head-Gordon08_084106} functionals. The optimized HF and DFT structures are provided in the Supporting Information. Subsequent QED-HF and QED-DFT calculations were carried out using the appropriate geometry and the same basis set. Each calculation was done using density-fitted two-body integrals, using the default JK-type auxiliary basis set for d-aug-cc-pVTZ in \textsc{Psi4} (def2-universal-JKFIT\cite{Weigend08_167}). The results reported in this work were obtained using the unrestricted formulations of HF and DFT.

{Lastly, we also examine the convergence of the cavity-induced changes in the static polarizability and first hyperpolarizability of $p$-nitroaniline using the cc-pV$n$Z, aug-cc-pV$n$Z, and d-aug-cc-pV$n$Z basis sets with $n=$ D, T, and Q. These calculations were performed using HF and at the geometry optimized at the HF/d-aug-cc-pVTZ level of theory mentioned earlier. Each calculation was carried out using the default JK-type auxiliary basis sets for cc-pV$n$Z, aug-cc-pV$n$Z, and d-aug-cc-pV$n$Z in \textsc{Psi4}, which are cc-pV$n$Z-JKFIT,\cite{Weigend02_4285} aug-cc-pV$n$Z-JKFIT,\cite{Weigend02_4285} and def2-universal-JKFIT,\cite{Weigend08_167} respectively.}

\section{Results and Discussion}

\label{SEC:RESULTS}

\begin{figure}[!htpb]
\includegraphics[width=0.9\columnwidth]{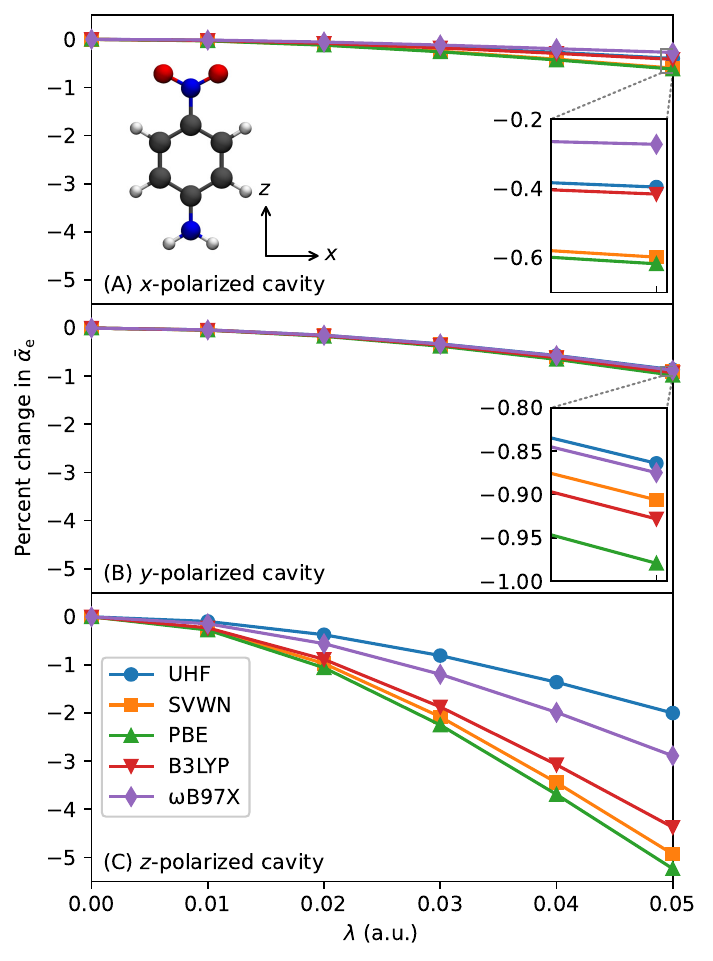}
        \caption{Cavity-induced changes to the static isotropically averaged electric dipole polarizability for $p$-nitroaniline computed using QED-HF and QED-DFT with different relative orientations of the molecule and cavity mode polarization axis. 
        The cavity frequency in the QED-HF/DFT calculations is set at 0.1 E$_\text{h}$ ($\approx 2.72$ eV).} 
        \label{FIG:ALPHA}
\end{figure}

We begin this discussion by assessing cavity induced changes to the isotropically averaged polarizability 
\begin{align}
    \bar{\alpha} = \frac{1}{3}\sum_{i} \alpha_{\text{e},ii};~i\in\{x,y,z\}
\end{align} 
for $p$-nitroaniline described at the QED-HF and QED-DFT levels. Figure \ref{FIG:ALPHA} depicts the percent change in $\bar{\alpha}$ when the molecule interacts with a single-mode cavity with a frequency of 0.1 E$_\text{h}$ ($\approx$ 2.72 eV) as a function of the coupling strength, $\lambda = |{\bm \lambda}|$, relative to the $\bar{\alpha}$ values computed for the isolated molecule using standard HF and DFT. This coupling strength is related to the effective mode volume, which, in atomic units, takes the form
\begin{align}
    \lambda = \sqrt{\frac{4\pi}{V_\text{eff}}}
\end{align}
The largest coupling strength we consider ($\lambda = 0.05$ a.u.) corresponds to an optical ``picocavity''\cite{Baumberg22_5859} with $V_\text{eff} \approx 0.74$ nm$^3$.
The molecule lies in the $xz$-plane, with the principal molecular axis aligned along the $z$-direction, and the data provided in panels (A), (B), and (C) of Fig.~\ref{FIG:ALPHA} correspond to the polarization axis being aligned along the $x$-, $y$-, and $z$-directions, respectively. In this configuration, the electronic and nuclear dipole moments point in the $z$ and $-z$ directions, respectively, and the total dipole moment points in the $-z$ direction.

When the cavity mode is polarized along the $x$-direction [Fig.~\ref{FIG:ALPHA}(A)], the isotropically averaged polarizability does not change much with increased coupling strength. Even at $\lambda = 0.05$ a.u., the largest changes to $\bar{\alpha}$ that we observe are only $\approx - 0.6$\% for the QED-DFT with the SVWN and PBE functionals, and the polarizability is even less sensitive to the presence of the cavity for QED-HF and the QED-DFT functionals that include exact exchange (either globally, with B3LYP, or with the range-separated $\omega$B97X functional). The situation is similar when the cavity mode is polarized in the $y$-direction [Fig.~\ref{FIG:ALPHA}(B)], where the largest change to $\bar{\alpha}$ is $\approx - 1$\% for QED-DFT with the PBE functional. Again, the inclusion of exact exchange can lessen the sensitivity of the polarizability to the cavity, but the pattern is slightly different than in panel (A); the sensitivity of the polarizabilities derived from QED-DFT with the SVWN and B3LYP functionals are reversed, as are those for QED-DFT with the $\omega$B97X functional and QED-HF. The isotropically averaged polarizability is most sensitive to the presence of the cavity when the cavity mode is polarized along the $z$-direction [Fig.~\ref{FIG:ALPHA}(C)], where the largest changes observed are $\approx -5$\% for QED-DFT with the SVWN and PBE functionals. Among the different polarizations we consider, the $z$-polarized cavity mode results in the greatest spread in the cavity-induced changes to $\bar{\alpha}$ for QED-HF and the various QED-DFT approximations. As in the other cases, the inclusion of exact exchange generally lessens the sensitivity of $\bar{\alpha}$ to the cavity, and we note that 100\% exact short- and long-range exchange, as is done in QED-HF, produces the least sensitive polarizability compared to the QED-DFT calculations.

\begin{figure}[!htpb]
\includegraphics[width=0.9\columnwidth]{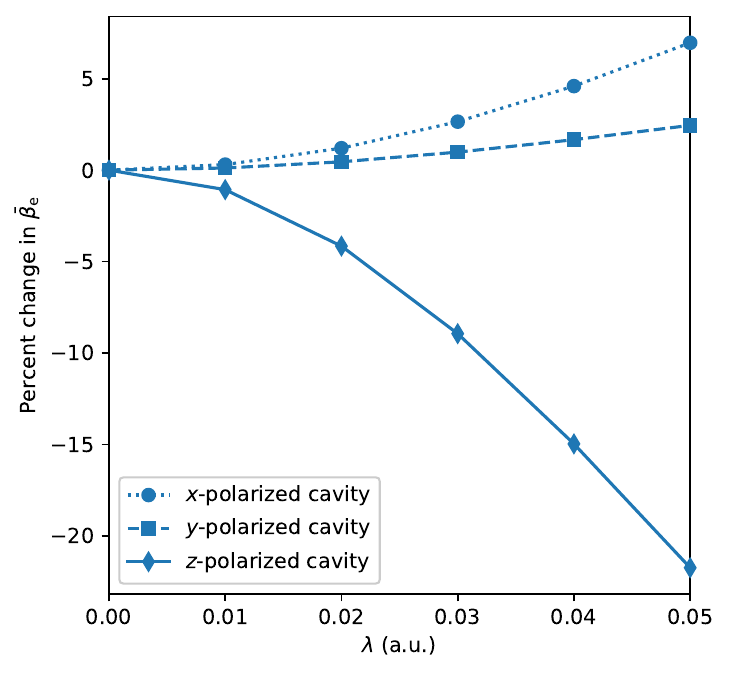}
        \caption{Cavity-induced changes to the static isotropically averaged first hyperpolarizability for $p$-nitroaniline computed using QED-HF with different relative orientations of the molecule and cavity mode polarization axis.
        The cavity frequency in the QED-HF/DFT calculations is set at 0.1 E$_\text{h}$ ($\approx 2.72$ eV).} 
        \label{FIG:BETA}
\end{figure}

We now consider cavity induced changes to the isotropically averaged first hyperpolarizability 
\begin{align}
\bar{\beta} = \frac{1}{5} \sum_{i}(\beta_{\text{e},iii} + \sum_{j\neq i}\beta_{\text{e},ijj});~i,j\in\{x,y,z\}
\end{align} 
for $p$-nitroaniline described at the QED-HF level. Figure \ref{FIG:BETA} depicts the percent change in $\bar{\beta}$ for the same single-mode cavity considered above, relative to the $\bar{\beta}$ value for the isolated molecule computed using standard HF. Unlike the polarizability, the hyperpolarizability can either be enhanced or suppressed via the cavity mode, depending on the relative orientation of the molecule and the polarization axis. For $x$- and $y$-polarized cavity modes, the isotropically averaged hyperpolarizability increases slightly with increase coupling strength, by as much as 7\% ($x$-polarized) and 2\% ($y$-polarized) at $\lambda = 0.05$ a.u. On the other hand, $\bar{\beta}$ decreases substantially when the cavity mode is polarized in the $z$-direction, by as much as $-$22\% (or from 133.5 a.u.~to 104.4 a.u.) at $\lambda = 0.05$ a.u. It is also worth noting that the first hyperpolarizability is most sensitive to $z$-polarized mode, which is similar to the pattern we observe in Fig.~\ref{FIG:ALPHA} for the polarizability. However, the change in $\bar{\beta}$ inside $y$-polarized cavity is smaller than that in the $x$-polarized one, which is the reverse of the relative sensitivity pattern observed for $\bar{\alpha}$.

Note that neither the static electric dipole polarizability nor the static first hyperpolarizability exhibit any  dependence on the cavity mode frequency, $\omega_\text{cav}$. This behavior is expected for mean-field cavity QED approaches formulated within the coherent-state basis where  the ground-state energy does not depend on $\omega_\text{cav}$ (see Eq.~\ref{EQN:QED_HF_ENERGY}). As for derivatives with respect to a static external electric field, it is easy to see that such properties will also be independent of $\omega_\text{cav}$ from the viewpoint of finite-field derivative calculations. If we apply the QED-HF or QED-DFT procedure to the perturbed Hamiltonian in Eq.~\ref{EQN:PERTURBED_H}, then the static electric field will modify the molecular electric dipole moment and, as a result, the coherent-state transformation operator (see Eq.~\ref{EQN:U_CS}). The total energy of the perturbed system will still have the form of Eq.~\ref{EQN:QED_HF_ENERGY}, with the modified dipole self energy term and an additional contribution from the external field. No dependence on $\omega_\text{cav}$ will be introduced. Numerical evidence for this behavior is provided in the Supporting Information, where we carry out frequency sweeps from $\omega_\text{cav}$ = 2--6 eV, for a $z$-polarized cavity with a coupling strength of $\lambda = 0.05$ a.u. As expected, neither the ground-state energies nor the static response properties are affected by the cavity frequency.

{ 
The} orientation dependence of the polarizability and hyperpolarizability deserve additional discussion. As can be seen in Fig.~\ref{FIG:ALPHA}, the isotropically averaged static electric dipole polarizability decreases with increased coupling strength for all cavity mode polarizations we have considered. The story is quite different for the static first hyperpolarizability, where we find that $\bar{\beta}$ is slightly enhanced when the molecule interacts with and $x$- or $y$-polarized cavity mode, whereas it decreases significantly when interacting with a cavity mode polarized in the $z$-direction. It is well known that cavity induced changes to electronic structure are highly dependent upon the relative orientations of the molecule and cavity mode polarization axis. It has also been argued that meaningful predictions of cavity modified properties should either involve appropriate orientational averaging \cite{Koch24_e1684,Koch24_8838} or account for cavity-induced rotations and geometric distortions.\cite{DePrince24_064109} From the geometry relaxation point of view, we note that the energy of $p$-nitroaniline is the lowest when the cavity mode is polarized along the $y$-axis, which happens to be the orientation associated with the second-smallest change to the polarizability and the smallest change to the hyperpolarizability. Hence, our ability to engineer large changes to response properties via cavity interactions depends on our ability to precisely control the relative orientations of the molecule and cavity mode polarization axis. 

\begin{figure}[!htpb]
\includegraphics[width=0.9\columnwidth]{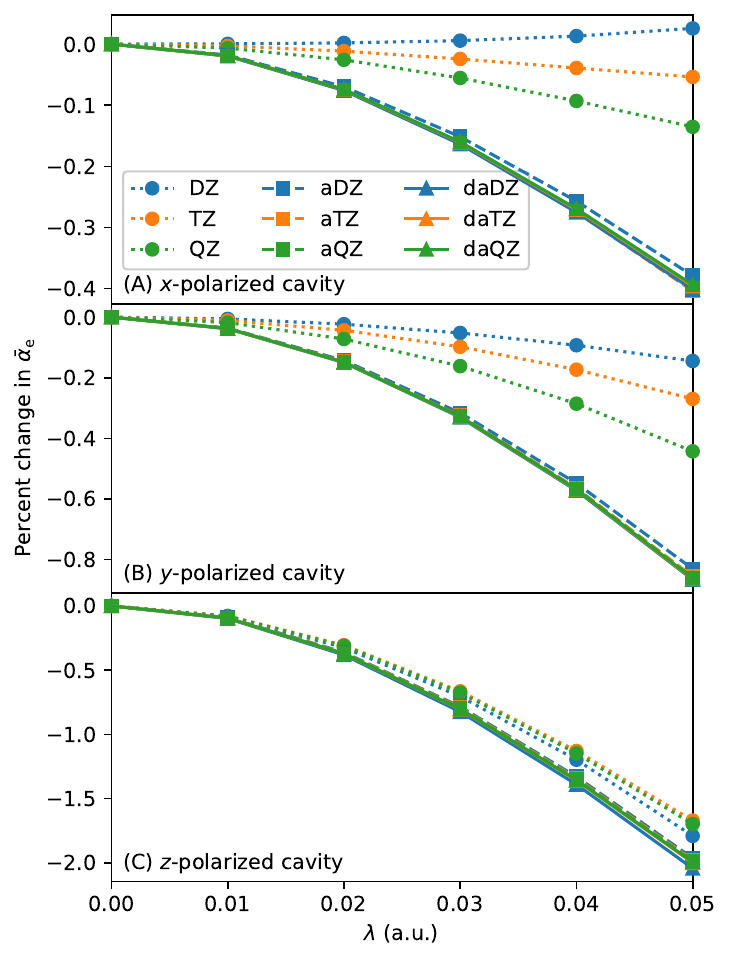}
        \caption{Cavity-induced changes to the static isotropically averaged electric dipole polarizability for $p$-nitroaniline computed using QED-HF with different relative orientations of the molecule and cavity mode polarization axis, as well as different basis sets ($n$Z = cc-pV$n$Z, a$n$Z = aug-cc-pV$n$Z, and da$n$Z = d-aug-cc-pV$n$Z). 
        The cavity frequency in the QED-HF calculations is set at 0.1 E$_\text{h}$ ($\approx 2.72$ eV).} 
        \label{FIG:ALPHA_BASIS}
\end{figure}

\begin{figure}[!htpb]
\includegraphics[width=0.9\columnwidth]{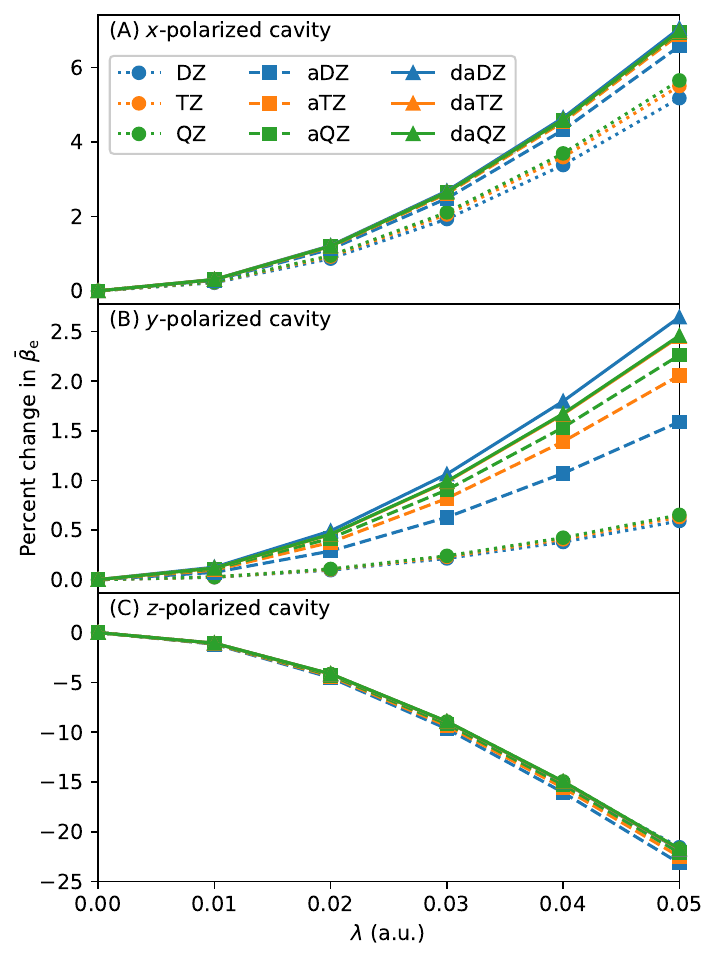}
        \caption{Cavity-induced changes to the static isotropically averaged first hyperpolarizability for $p$-nitroaniline computed using QED-HF with different relative orientations of the molecule and cavity mode polarization axis, as well as different basis sets ($n$Z = cc-pV$n$Z, a$n$Z = aug-cc-pV$n$Z, and da$n$Z = d-aug-cc-pV$n$Z). 
        The cavity frequency in the QED-HF calculations is set at 0.1 E$_\text{h}$ ($\approx 2.72$ eV).} 
        \label{FIG:BETA_BASIS}
\end{figure}

{Lastly, we consider the convergence of the cavity-induced changes in the response properties considered above as a function of the basis set size. Figure~\ref{FIG:ALPHA_BASIS}  illustrates the percent change in the isotropically averaged static electric dipole polarizability as a function of the coupling strength, when the calculations are carried out using the cc-pV$n$Z, aug-cc-pV$n$Z, and d-aug-cc-pV$n$Z basis sets (with $n=$ D, T, and Q).
Cavity-induced changes to the $\bar{\alpha}_{\text e}$ values for $p$-nitroaniline are essentially converged at at any $\zeta$-level, provided that one uses augmented (aug-cc-pV$n$Z) or doubly-augmented (d-aug-cc-pV$n$Z) basis sets. On the other hand, the percent changes in $\bar{\alpha}_{\text e}$ obtained using the cc-pV$n$Z basis sets differ significantly from results obtained from (doubly-) augmented basis sets and also display a clear dependence on the $\zeta$-level. At $\lambda$ = 0.05 a.u., the percent change in $\bar{\alpha}_{\text e}$ predicted in these basis sets shows a $\approx$ 0.2\% spread, and these values themselves differ by 0.2\%--0.4\% from the converged d-aug-cc-pV$n$Z results. These observations are similar to those reported for the smaller LiH, H$_2$O, and CO molecules in Ref.~\citenum{Kowalewski25_5171}.  Note, also, that calculations carried out in the cc-pVDZ basis set predict that the magnitude of the isotropically averaged polarizability increases with coupling strength when the cavity is polarized in the $x$-direction, which is qualitatively incorrect, relative to results obtained from calculations that use basis sets with diffuse functions. 

Figure~\ref{FIG:BETA_BASIS}  illustrates the percent change in the isotropically averaged static first hyperpolarizability as a function of the coupling strength, when the calculations are carried out using the same basis sets considered in Fig.~\ref{FIG:ALPHA_BASIS}. As in Fig.~\ref{FIG:ALPHA_BASIS}, the predicted changes in $\bar{\beta}_{\text e}$ obtained from calculations carried out in non-augmented basis sets are significantly different from those obtained from calculations that use doubly-augmented basis sets; at $\lambda=0.05$ a.u., cc-pV$n$Z-derived results differ from those obtained using the d-aug-cc-pVQZ basis by roughly 1\%--2\%, with the results from the z-polarized cavity having the smallest discrepancies. Unlike in Fig.~\ref{FIG:ALPHA_BASIS}, we find that results obtained using aug-cc-pV$n$Z basis sets display a non-trivial $\zeta$-level dependence [see Fig.~\ref{FIG:BETA_BASIS} (B), in particular]. The data included in the Supporting Information indicate that, in order to obtain $\bar{\beta}_{\text e}$ values that agree to within $\approx$1\% with those obtained from calculations performed using the largest basis set considered in this work (d-aug-cc-pVQZ), one should use at least an aug-cc-pVTZ basis set.}


\section{Conclusions}

\label{SEC:CONCLUSIONS}

A large number of recent computational studies have explored the influence of strong electron--photon coupling on the electronic structure of cavity-embedded molecular systems. Many such studies have focused on the impact that strong coupling can have on ground-state properties, with comparatively fewer works focusing on molecular response properties, particularly higher-than-first-order responses. In this work, we have developed mean-field approaches to model the response of a molecule within an optical cavity to static external electric fields. While the electric dipole polarizability (a first-order response property) is not particularly sensitive to the presence of the cavity, the first hyperpolarizability (a second-order property) can be significantly impacted by the cavity, depending on the cavity mode polarization axis.
{Our calculations also indicate that not only are first hyperpolarizabilities more sensitive to the basis set size and composition when compared to polarizabilities, which is well known (see, e.g., Ref.~\citenum{Maroulis85_2380,Bartlett93_3022,Maroulis98_5432,Masunov08_044109}), but the cavity-induced changes in $\bar{\beta}_{\text e}$ are also more affected by the basis set choice than the corresponding changes in $\bar{\alpha}_{\text e}$. Interestingly, even though the static polarizabilities are less sensitive to basis sets than the hyperpolarizabilities, the use of small basis set and the lack of diffuse functions can still lead to qualitatively incorrect predictions of cavity-induced changes to $\bar{\alpha}_{\text e}$ as shown in Fig.~\ref{FIG:ALPHA_BASIS} (A).}
Moving forward, it will be interesting to examine cavity-induced changes to higher-order static response properties, as well as their dynamic analogues, in which case we expect the cavity mode frequency to play an important part. Lastly, it also will be important to consider how explicit electron--photon correlations that are captured by many-body {\em ab initio} QED approaches impact the response properties of cavity-embedded systems, as it has repeatedly been demonstrated that mean-field approaches such as those studied herein tend to exaggerate cavity-induced changes to electronic structure.\cite{DePrince21_094112, DePrince23_5264, DePrince24_064109} 

\vspace{0.5cm}

\noindent {\bf Supporting Information}

The cartesian coordinates for $p$-nitroaniline obtained from geometry optimizations using HF, SVWN, PBE, B3LYP, and $\omega$B97X and the d-aug-cc-pVTZ basis set, the ground-state energies and response properties of $p$-nitroaniline computed using QED-HF and QED-DFT at different $\lambda$ values and a fixed cavity frequency of 0.1 E$_\text{h}$, the excitation energies and oscillator strenghts of the first bright $A_1$ state of $p$-nitroaniline obtained using QED-HF and QED-DFT, {the} ground-state energies and response properties of $p$-nitroaniline computed using QED-HF and QED-DFT at $\lambda = 0.05$ a.u.~and different cavity frequencies{, and the response properties of $p$-nitroaniline computed using HF at different $\lambda$ values and basis sets with a fixed cavity frequency of 0.1 E$_\text{h}$.}

\vspace{0.5cm}

\noindent {\bf Acknowledgements}

This material is based upon work supported by the National Science Foundation under Grant No. CHE-2100984.

\vspace{0.5cm}



\bibliography{Journal_Short_Name,main}

\end{document}